\documentclass[12pt]{article}
\usepackage{sc3conf}
\usepackage{amsfonts}
\usepackage{epsfig}

\begin{document}
\raggedbottom \noindent\hfill\hbox{\rm  } \vskip 1pt
\noindent\hfill\vbox{\hbox{\rm SLAC-PUB-9595} \hbox{\rm November
2002}} \vskip 30pt
\title{
%
%
\[ \vspace{-2cm} \]
%
Black Holes, Hawking Radiation and the Information
Paradox\footnotemark[1]} \footnotetext[1]{This research was
supported by the DOE under grant number DE-AC03-76SF00515}
\authors{Marvin Weinstein}

\addresses{Stanford Linear Accelerator Center\\
Stanford University, Stanford, CA 94309\\
E-mail: niv@slac.stanford.edu}

\maketitle

\begin{abstract}
  This talk is about results obtained by Kirill Melnikov and myself
  pertaining to the canonical quantization of a massless scalar field
  in the presence of a Schwarzschild black hole.  After a brief
  summary of what we did and how we reproduce the familiar Hawking
  temperature and energy flux, I focus attention on how our
  discussion differs from other treatments.  In particular I
  show that we can define a system which fakes an
  equilibrium thermodynamic object whose entropy is given by the
  $A/4$ (where $A$ is the area of the black hole horizon), but for
  which the assignment of a classical entropy to the system
  is incorrect.  Finally I briefly discuss a discretized version
  of the theory which seems to indicate that things work in a
  surprising way near $r=0$.
\end{abstract}

\newcommand{\ba}{\begin{eqnarray}}
\newcommand{\ea}{\end{eqnarray}}
\newcommand{\p}{\mbox{$\vec{p}$}}
\newcommand{\e}{\mbox{$\vec{e}$}}
\newcommand{\n}{\mbox{$\vec{n}$}}
\newcommand{\vsig}{\mbox{$\vec{\sigma}$}}
\newcommand{\be}{\begin{equation}}
\newcommand{\ee}{\end{equation}}
\def\eq#1{(\ref{#1})}
\newcommand{\bra}[1]{\langle  {#1}  \vert }
\newcommand{\ket}[1]{\vert {#1} \rangle }

\section{Introduction}

In this talk I will discuss results obtained by Kirill Melnikov
and myself\cite{melwein} pertaining to the canonical quantization
of a massless scalar field theory in the presence of a
Schwarzschild black hole.  The main difference of this work from
earlier work is that we study the future evolution of a well
defined state of a quantum field theory defined on a given
space-like hypersurface as a function of time and show that we can
obtain explicit expressions for Hawking radiation\cite{hawking}
without computing things at null infinity.

In particular, we show that for a large black hole, the usual
formula for Hawking radiation is obtained well before one is
forced to deal with the evaporation of the hole and within a
Hamiltonian framework, which explicitly preserves unitarity. We
actually derive a variant of the usual Hawking result: namely, if
one starts from any reasonable state and waits long enough, then
an Unruh thermometer\cite{unruh} located at a fixed Schwarzshild
$r$ will measure a temperature inversely proportional to the mass
of the black hole.  We also show there is a uniquely defined
finite energy flux through any sphere of large but finite
Schwarzschild $r$ and any large but finite time, $t$, and that the
size of this flux agrees with the Hawking result.

Since we derive the usual Hawking results for the apparent
temperature of a black hole it is fair to ask what we have to say
that is new.  To our minds the most important difference is that
our derivations are done within an explicitly unitary framework
and we derive all results for observers located at a finite
distance from the black hole and for large but finite times after
the initial state begins to evolve. This differs significantly
from derivations which compute transitions from past null infinity
to future null infinity.  Furthermore, this difference allows us
to construct a variation of the original problem which shows that
an observer could mistakenly conclude he is dealing with an
equilibrium system with a Bekenstein entropy\cite{bekenstein} well
before any question of black hole evaporation or information loss
becomes relevant.  Needless to say, this leads to a difference in
interpretation which I will discuss at the appropriate time.

The physical picture which emerges from our analysis is that,
consistent with the general theorem which says that the
Scwarzschild metric does not admit any global timelike Killing
vector field, our globally defined Hamiltonian is perforce time
dependent. It is this time dependence which provides the mechanism
which generates the Hawking radiation. Moreover, since this time
dependence continues forever, we see that the Hawking radiation is
not to be viewed as an equilibrium phenomenon, but rather as the
steady state behavior which one might expect in such a system.
There is no finite time at which the concepts of equilibrium
thermodynamics apply.

\section{The Plan}

I begin by spending a few moments showing you how the foliation of
Schwarzschild space-time is done and then briefly discuss the
quantization of the scalar field theory. After that I briefly show
how the Heisenberg equations of motion are used to study the time
evolution of the system and how these are approximately solved in
the Schwarzschild background metric.  With these preliminaries out
of the way I focus on variants of the problem which address
physics issues which come up.  First, I address the question of
how to choose an initial state.  After this I exhibit a variant of
the problem which shows how an outside observer can conclude that
he is studying an equilibrium system with a well defined
temperature and entropy, even though nothing could be further from
the truth.  Finally, I conclude by discussing the much more
interesting question of what is happening near the real
singularity at $r=0$. The ideas in this part of my talk are much
more speculative, but the results are thought provoking.

\section{Preliminaries}

I begin by considering a massless scalar field theory with
Lagrangian density
\begin{equation}
    {\cal L} = \sqrt{-g} \left[ g^{\mu \nu} \,\partial_\mu \phi(x)\,
     \partial_\nu \phi(x) \right]
\label{masslesslag}
\end{equation}
in the background of a Schwarzschild black hole of mass $M$.  In
the usual Schwarzschild coordinates the metric $g_{\mu \nu}$ takes
the familiar form
\begin{equation}
    ds^2 =
- (1-{2M\over r} )\,dt^2 +  (1- {2M\over r} )^{-1}\,dr^2 + r^2
d\Omega^2 , \label{schwarz}
\end{equation}
where we have set Newton's constant, $G$, to one. As is well
known, the apparent metric singularity at $r=2M$ is a coordinate
artifact and, as such, does not pose a problem. The true issue for
canonical quantization is that we need to define a family of
spacelike slices which foliate the spacetime in order to define
initial data and form the Hamiltonian. Inspection of
Eq.(\ref{schwarz}) shows that surfaces of constant Schwarzschild
time change from spacelike to timelike at the horizon ($r=2M$) and
so they do not fulfill our requirements.

\begin{figure}[htb]
   \begin{center}
    \epsfig{file=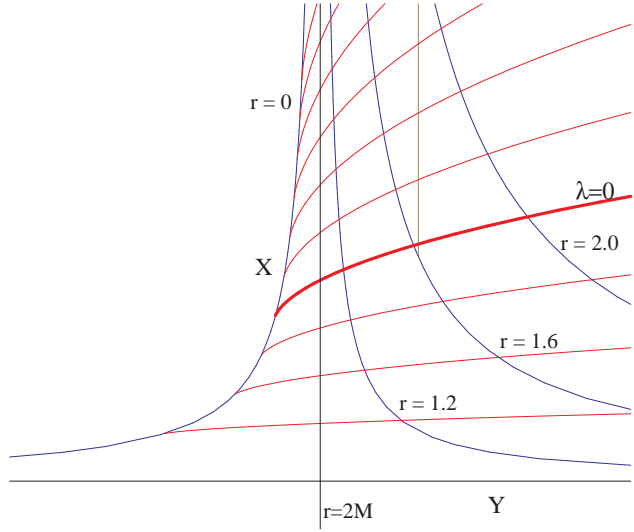,width=3.25in}
     \caption{Space-time foliation overlaid on Kruskal coordinates.}\label{kruskal}
   \end{center}
 \end{figure}

In order to exhibit a satisfactory family of spacelike slices it
is convenient to introduce dimensionless versions of $r$ and $t$
by rescaling $ r \rightarrow 2Mr $ and $ t \rightarrow 2M t $.
Using these variables we introduce the Kruskal coordinates $X$ and
$Y$ by the equations:
\begin{equation}
 X Y = (r-1)\,e^{r}, \qquad  \qquad
        {X \over \vert Y \vert} = e^{t_S} .
\label{kruskaleq}
\end{equation}
In these coordinates the Schwarzschild metric takes the form
\begin{equation}
 ds^2 = {32\,  e^{\vert -r \vert} dX\, dY \over r} + r^2 d\Omega^2 .
\end{equation}
Eq.(\ref{kruskaleq}) tells us that fixed Schwarzschild $r$ is a
hyperbola in the $X,Y$-plane, as shown in Fig.\ref{kruskal}, and
that a surface of fixed Schwarzschild time corresponds to a
straight line $X=\vert Y \vert\,e^{t_S}$ (such lines are not shown
in Fig.\ref{kruskal}).

Next we introduce Painlev\'e coordinates, which are derived from
Schwarzschild coordinates by making an $r$-dependent shift in
Schwarzschild time, i.e.,
\begin{equation}
 t = \lambda - 2\sqrt{r} - \ln\left(\left\vert{\sqrt{r}
- 1 \over \sqrt{r} + 1}\right\vert\right).
\end{equation}
The spacelike surfaces we wish to define correspond to surfaces of
fixed Painlev\'e time.  They are the almost horizontal curves
shown in Fig.\ref{kruskal}.  It is obvious these surfaces are
everywhere spacelike since, in Painlev\'e coordinates, the
Schwarzschild metric takes the form
\begin{equation}
ds^2 = - \left (1 - {1 \over r} \right ) \,d\lambda^2 +
{2\,d\lambda\,dr \over \sqrt{r}} + dr^2 +  r^2 d\Omega^2 .
\label{painlevemetric}
\end{equation}

To carry out the canonical quantization of the scalar field theory
it is convenient to make one more change of variables. This leads
to Lema\^itre coordinates, which are related to Painlev\'e
$\lambda$ and $r$ by
\begin{equation}
    r(\lambda,r_{sch}) = \left (
r_{sch}^{3/2} - {3\over 2}\lambda \right )^{2/3}
    = \left ( {3\over 2} (\eta - \lambda) \right )^{2/3}.
\end{equation}
In Lema\^itre coordinates the metric takes the form
\begin{equation}
ds^2 = - d\lambda^2 + {1\over r(\lambda,\eta)} d\eta^2 +
r(\lambda,\eta)^2 d\Omega^2.
\end{equation}

The Lema\^itre form of the metric has the property that it is
manifestly free of coordinate singularities at $r=1$, has no cross
terms in $d\lambda$ and $dr$, and allows a completely
straightforward canonical quantization procedure.

\section{Canonical Quantization}

The Lagrangian for the massless scalar field in Lema\^itre
coordinates is rotationally invariant, so we can study the theory
one angular momentum mode at a time. Expanding the field
$\phi(\lambda, \eta, \theta, \phi)$ in spherical harmonics in
$\theta$ and $\phi$ and restricting attention to the $L=0$ mode we
find the Lema\^itre coordinate form of the $L=0$ scalar field
Lagrangian to be
\begin{equation}
{\cal L} = \sqrt{-g} \,{1\over 2} \left[ (\partial_\lambda
\phi_0\,(\lambda,\eta))^2 -
r\,(\partial_\eta\phi_0\,(\lambda,\eta))^2 \right]
\end{equation}
where the determinant $\sqrt{-g}$ is
\begin{equation}
\sqrt{-g} = r^{3/2} = {3\over 2} (\eta-\lambda).
\end{equation}

From this we see that the momentum conjugate to the field is
\begin{equation}
\pi_0(\lambda,\eta) =  {3\,(\eta - \lambda)\over
2}\,\partial_{\lambda}\phi_0(\lambda,\eta ) ,
\end{equation}
and the canonical Hamiltonian is
\begin{equation}
H(\lambda) = {1\over 2} \int_{\lambda}^\infty d\eta\, \left(
{2\,\pi_0(\lambda,\eta)^2 \over 3(\eta - \lambda)} + {3\over 2}
r\,(\eta-\lambda) (\partial_\eta\phi_0(\lambda,\eta))^2 \right) .
\label{hamone}
\end{equation}
The commutation relations for $\phi_0$ and $\pi_0$ are
\begin{equation}
\left[ \pi_0(\lambda,\eta), \phi_0(\lambda,\eta') \right] =
-i\,\delta(\eta - \eta').
\end{equation}

It follows from Eq.\ref{hamone}  and the canonical commutation
relations that the Heisenberg equation of motion for the field is
\begin{equation}
\label{scheulerlag}
\partial_{\lambda} \left[ (\eta - \lambda) \partial_{\lambda}\phi_0 \right]
-\partial_{\eta}\left[ (\eta - \lambda)\,r\,\partial_{\eta}\phi_0
\right] = 0.
\end{equation}

Before discussing the solution of the Heisenberg equations of
motion I want to emphasize that it is simple to find all of the
eigenstates of $H(0)$, because it is just a free field Hamiltonian
in disguise.  To see this change variables back to Schwarzschild
$r$, using $\eta=(2/3) r^{3/2}$ and rescale the fields by
\begin{eqnarray}
    \pi_0(r) = \sqrt{r}\,\pi_1(r),~~~
\phi_0(r) = {\phi_1(r)\over r} .
\end{eqnarray}
This converts Eq.(\ref{hamone}) to
\begin{equation}
H(0) = {1\over 2} \int_{0}^\infty dr\, \left( \pi_1(r)^2  +
r^2(\partial_r\,{\phi_1\over r})^2 \right) , \label{hamtwo}
\end{equation}
which  is the Hamiltonian of the $L=0$ mode of a free massless
field in flat space.  To construct the eigenstates of this
Hamiltonian simply expand the fields in terms of annihilation and
creation operators,
\begin{eqnarray}
    \phi_1(r) =
\int_0^{\infty} {d \omega \over \sqrt{\pi\,\omega }}\,\sin(\omega
r)
    \left( a^{\dag}_\omega + a_\omega \right),~~~
 \pi_1(r) =
i \int_0^{\infty} d\omega \,\sqrt{ \omega \over \pi}\,\sin(\omega
r)
    \left( a^{\dag}_\omega - a_\omega \right) ,
\label{ffexpansion}
\end{eqnarray}
and define the vacuum state $\ket{0}$ to be the state that is
annihilated by all of the $a_\omega$'s.

\section{Geometric Optics Approximation}

To see how the initial state evolves it is best to use the
Heisenberg representation and solve for the Heisenberg fields at
later times as a function of the fields defined on the initial
surface of quantization.  To do this we use a geometric optics
approximation.

To briefly describe this geometric optics approximation let us
study Eq.(\ref{scheulerlag}) in Painlev\'e coordinates
($\lambda,r$), since these coordinates are non-singular and the
dependence of the solutions on $\lambda$ and $r$ factorizes. The
WKB approach to this problem is to assume a solution of the form
$\phi_0 = r^{-1} e^{i \omega \lambda} f_\omega(r)$ and substitute
this ansatz into the field equation.  In this way one obtains
that, for large $\omega$, $f_\omega(r)$ can be written  as
\begin{eqnarray}
\ln f_\omega(r) &=& i\omega S_{1,2}(r) + {\cal O}(\omega^{-1}), ~
S_{1,2}(r) = \pm r -2\sqrt{r} \pm \ln((\sqrt{r}\pm 1)^2).\nonumber\\
\label{solutions}
\end{eqnarray}
We now observe that these solutions are constant along incoming or
outgoing null geodesics where an incoming null-geodesic starting
at the point $x_1$ at time $\lambda=0$ is a curve $r(\lambda)$
such that
\begin{equation}
    S_1(x_1) = \lambda + S_1(r(\lambda)),
\label{ngone}
\end{equation}
and similarly, an outgoing geodesic starting at $x_2$ at
$\lambda=0$, is a curve $r(\lambda)$ such that
\begin{equation}
    S_2(x_2) =  \lambda + S_2(r(\lambda)),
\label{ngtwo}
\end{equation}
where $S_{1,2}$ are as defined in Eq.(\ref{solutions}).

It is simple to convert this observation into an ansatz for the
solution to the S-wave field equation by imitating the general
form of the solution of the same sort of problem in flat space;
i.e. we say that for general $(\lambda,r)$
\begin{eqnarray}
    \phi_0(\lambda,r) &=& {1 \over r} \left( \tilde\phi_1 (\lambda + S_1(r))
    + \tilde\phi_2 ( \lambda+S_2(r) ) \right),
\label{geomopt}
\end{eqnarray}
and the functions $\tilde \phi_{1,2}(S_{1,2}(r))=f_{1,2}(r)$ are
to be determined from the boundary conditions
\begin{equation}
    \phi_0(0,r) = {\phi_1(r)\over r}, \quad
    \partial_\lambda \,\phi(\lambda,r)\vert_{\lambda=0} = \sqrt{r}\pi_1(r),
\label{bndrycond}
\end{equation}
where $\phi_1(r)$ and $\pi_1(r)$ are the rescaled operators we
introduced to quantize the theory on the initial surface
$\lambda=0$.

Substituting Eq.(\ref{geomopt}) into Eq.(\ref{bndrycond})  we
obtain \be f_{1,2}(x) = \frac {1}{2} \int \limits_{0}^{x} {\rm d}
\xi \left [ \phi_1'(\xi) \pm \pi_1(\xi) \mp \frac{
\phi_1(\xi)}{\xi^{3/2}} \right ], \label{funcf} \ee where
$\phi_1'= {\rm d}\phi_1/{\rm d}\xi$ and $S_{1,2}(x_{1,2}) =
\lambda + S_{1,2}(r)$. Given that the field $\phi_1$ and its
momentum $\pi_1$ are expressed through the creation and
annihilation operators defined at $\lambda=0$, we can compute any
Green's function of the field $\phi_0$ at any later time.

I will refer you to our paper\cite{melwein} to see how the
calculation for the response of the Unruh thermometer and outgoing
flux are carried out.  The important point of these calculations
are that the effect gets its contributions from null geodesics
which leave the initial surface of quantization from points just
outside, but exponentially close to, the horizon.

\section{The Infalling Mirror}

Having described the general framework, a question comes to mind;
namely, "Is there a problem because our initial state is defined
to be a single coherent quantum state both inside and outside the
horizon?".  The issue is that there is no physical mechanism for
preparing such a system, since there is no way degrees of freedom
inside the horizon communicate with those outside.  To address
this issue and show it is a non-problem we considered a variant of
the original problem in which we imagine a black hole which, up to
some Lema\^itre time $\lambda$, is surrounded by a perfectly
reflecting mirror of radius $R_0$.  This is technically
implemented by assuming the field vanishes inside and on the
spherical surface $R_0$ at large times in the past. In this way we
guarantee that the physical state which we start from when we
quantize the theory is completely outside of the horizon.

Next, at some finite time $t$ we assume the mirror starts to fall
into the black hole along one of the Lema\^itre time-lines.  We
then carry out the computation for an Unruh thermometer, or
outgoing flux, at large times in the future and at large distances
from the black hole.  The result is, of course, unchanged. Details
do differ, however.  Now the null geodesics which arrive at the
measuring apparatus at late times do not originate from a point on
the initial surface of quantization at a point exponentially close
to the horizon (since there is no field inside the reflecting
sphere).  Rather, they come from geodesics which begin life as
infalling null geodesics which are then reflected from the
infalling sphere just before it crosses the horizon.  In this way
this problem behaves in much the same way as the analysis of a
black hole which is assembled by infalling radiation.

This variant of the problem actually establishes two facts: first,
that there is no real problem associated with starting from a
coherent quantum state defined to be inside and outside the
horizon; second, it shows that the point on the initial surface of
quantization which corresponds to the place from which the Hawking
radiation arises depends upon when one chooses to let go of the
mirror.  Clearly, in such a situation, modifying the initial state
so as to suppress the Hawking radiation is a very unphysical thing
to do.

\section{The Two Mirror Problem -- Bekenstein Entropy}

The next item I wish to discuss is a modification of the original
problem in which we place one mirror close to the black hole and
another at a large distance from the black hole. In this case, so
long as both mirrors remain static the theory has a well defined,
time-independent Hamiltonian and a unique ground state. So long as
both mirrors remain static an observer outside the larger mirror
will, by dropping a test charge, be able to measure the mass of
the black hole but nothing else.  If he sticks an Unruh
thermometer through a small hole in the surface he will measure
zero temperature.

Now, if we allow the inner mirror to collapse, as in the previous
section, then simple modification of the preceding analysis leads
to the following results: first, since all Hawking radiation is
reflected from the outer mirror, the rate of evaporation of the
black hole is proportional to $1/M^5$; second, at times which are
long, but short with respect to the time needed for significant
evaporation of the black hole, the outside observer who sticks an
Unruh thermometer through a small hole in the mirror will measure
a temperature proportional to $1/8 \pi M$; third, since no
radiation leaves the outer sphere the outside observer always
measures the same total mass, $M$.  Thus, at times long after the
inner mirror collapses the outside observer thinks he is dealing
with a body of energy $M$ and temperature $T_H = 1/8 \pi M$.
Following Bekenstein he would say that he is dealing with an
equilibrium thermodynamic system (since he sees nothing changing)
for which
\begin{equation}
    dU = dM = T dS = dS/8\pi M
\end{equation}
from which it follows that $ S = A_{BH}/4\pi$, where $A_{BH}$ is
the area of the horizon of the black hole.  However, he would
reach this conclusion for a system which is always in a pure
quantum state.  What is wrong with this analysis?

The answer is clear if we look inside the static mirror.  In that
case we realize that we are not dealing with an equilibrium system
at all.  Rather we have a system with a time dependent Hamiltonian
and the temperature we see is the result of steady-state and not
thermal behavior.  Nevertheless, from the point of view of the
outside observer there is no way to know this fact.

\section{What Is Going On At $r=0$?}

I must begin by emphasizing what should have been clear from the
rest of this paper, that none of what I have said addresses issues
associated with quantum gravity. Our work focuses on what is
happening if one studies field theory in a classical gravitational
background.  Having said this, there is an issue which we discuss
in our paper which overlaps with this question; namely, what is
going on near the real singularity at $r=0$.  Since, except for
the case of a two-dimensional black hole, the geometric optics
approximation doesn't work in this region we took a different
approach to analyze the problem; namely, we introduced a lattice
in the Lema\^itre coordinate $\eta$.  This lattice is peculiar in
that it does nothing to regulate the ultraviolet behavior of the
field theory at large distances but it does make things better
behaved inside the horizon near $r=0$.  Adopting the point of view
that too close to $r=0$ quantum gravity becomes important we
discuss a system for which the singularity is excised and a
surface put at some $r=\epsilon$.  We find, consistent with what
happens in the case of the two-dimensional black hole, that this
space-like surface must be included when one integrates to get the
Hamiltonian at times later than the time assigned to the surface
of quantization.  When we do this and solve for the eigenstates of
the Hamiltonian at different times we are surprised to find that
states which one would have thought to be totally contained on the
surface $r=\epsilon$ stretch through the horizon. It would take
too much space to discuss this here, but clearly this result
raises many questions which deserve further study.

\section{Conclusions}

To my mind there are three issues raised in this analysis. First,
despite what has been said by many, the behavior of a massless
quantum field in the background of a Schwarzschild black hole
seems to be unitary and no important issues arise, except for the
behavior of the theory near $r=0$ where we expect issues of
quantum gravity to significantly modify any semi-classical
analysis.  Thus, since no obstruction exists to treating this
system according to the usual rules of quantum mechanics I do not
believe any analysis of the semi-classical system will shed light
upon the question of what the true quantum completion of gravity
should be.  Second, the discussion of the two mirror example
raises a serious question of interpretation of the Bekenstein's
discussion of black hole entropy for the case of a Schwarzschild
black hole.  This discussion, which closely parallels Bekenstein's
original argument, shows that for the case of the Schwarzschild
black hole one can construct a steady-state system which, from the
point of view of an external observer, mimics the behavior of a
system in thermodynamic equilibrium without being one.  Finally,
the discussion of the discretized system raises the intriguing
possibility that due to the mixing of low-energy states, there is
something non-trivial to understand about how the quantum system
behaves with respect to information stored in the black hole.


\begin{thebibliography}{9}

\bibitem{melwein} ON UNITARY EVOLUTION OF A MASSLESS SCALAR FIELD IN A SCHWARZSCHILD
BACKGROUND: HAWKING RADIATION AND THE INFORMATION PARADOX. By
Kirill Melnikov, Marvin Weinstein (SLAC). SLAC-PUB-9224, May 2002.
45pp. e-Print Archive: hep-th/0205223


\bibitem{hawking}
S. W. Hawking, Commun. Math. Phys. {\bf 43}, 199 (1975); J.B.
Hartle and S.W. Hawking, Phys. Rev. {\bf D13}, 2188 (1976).


\bibitem{unruh} W.G. Unruh, Phys. Rev. {\bf D14}, 870 (1976).

\bibitem{bekenstein}
J.~D.~Bekensten, Phys. Rev. {\bf D7}, 2333 (1973),
J.~D.~Bekenstein, Phys. Rev. {\bf D9}, 3292 (1974).

\end{thebibliography}
\end{document}